\documentclass[twocolumn,aps,showpacs,superscriptaddress]{revtex4-1}
\usepackage{amsmath}
\usepackage{graphicx}
\RequirePackage[pagewise,mathlines]{lineno}

\begin{document}
\title{Coherent lepton pair production in hadronic heavy ion collisions}%
\author{W. Zha}\affiliation{University of Science and Technology of China, Hefei, China}
\author{L. Ruan}\affiliation{Brookhaven National Laboratory, New York, USA}
\author{Z. Tang}\email{zbtang@ustc.edu.cn}\affiliation{University of Science and Technology of China, Hefei, China}
\author{Z. Xu}\affiliation{Brookhaven National Laboratory, New York, USA}\affiliation{Shandong University, Jinan, China}
\author{S. Yang}\affiliation{Brookhaven National Laboratory, New York, USA}
\date{\today}%
\begin{abstract}
Recently, significant enhancements of $e^{+} e^{-}$ pair production at very low transverse momentum ($p_{T} < 0.15$ GeV/c) were observed by the STAR collaboration in peripheral hadronic A+A collisions. This excesses can not be described by the QGP thermal radiation and $\rho$ in-medium broadening calculations. This is a sign of coherent photon-photon interactions, which were conventionally studied only in ultra-peripheral collisions. In this article, we present calculations of lepton pair ($e^{+}e^{-}$ and $\mu^{+}\mu^{-}$) production from coherent photon-photon interactions in hadronic A+A collisions at RHIC and LHC energies within the STAR and ALICE acceptance.
\end{abstract}
\maketitle
The major scientific goal of relativistic heavy-ion collisions carried out at the Relativistic Heavy Ion Collider (RHIC) and the Large Hadron Collider (LHC) is to study the properties of the deconfined state of partonic matter - Quark-Gluon Plasma (QGP)~\cite{PBM_QGP}. Dileptons has been suggested as "penetrating probes" for the hot and  dense medium created in heavy-ion collisions~\cite{SHURYAK198071}, because they are produced during the whole evolution and not subject to the violent strong interactions in the medium. Various dilepton measurements have been performed since the early days of heavy-ion collisions~\cite{PhysRevLett.79.1229,PhysRevLett.98.052302,PhysRevC.84.014902,Angelisetal.2000,PhysRevLett.91.042301,PhysRevLett.96.162302,PhysRevLett.100.022302,Arnaldi2009,PhysRevC.81.034911,PhysRevLett.113.022301,Adamczyk201564,KOHLER2014665}. Of particular interest, a clear enhancement in the low ($M_{ll} < M_{\phi}$) and intermediate mass ($M_{\phi} < M_{ll} < M_{J/\psi}$) region~\cite{0954-3899-34-8-S149} has been observed when compared to known hadronic source. The enhancement in the low mass region is consistent with in-medium broadening of the $\rho$ mass spectrum~\cite{PhysRevLett.97.102301,VANHEES2008339,PhysRevLett.100.162301,PhysRevC.77.024907,PhysRevC.75.024908,PhysRevC.84.054917}, while the excess in the intermediate mass region is believed to be originated from the QGP thermal radiation~\cite{PhysRevC.63.054907}.

Dileptons can also be produced by the collision of the two intense electromagnetic fields which accompany the relativistic heavy ions~\cite{BAUR20071}. The electromagnetic field of a relativistic charge particle can be viewed as a spectrum of equivalent photons. The photon flux is proportional to $Z^{2}$, where $Z$ is the charge of the particle. The dilepton production from the electromagnetic interactions can be represented as $\gamma + \gamma \rightarrow l^{+} + l^{-} $. The equivalent two-photon luminosity is proportional to $Z^{4}$. The strong dependence on $Z$ results in copiously produced dileptons in relativistic heavy-ion collisions. Conventionally, the two-photon process was only studied without background from hadronic processes, $i.e.$ in the so-called Ultra-Peripheral Collisions (UPCs)~\cite{UPCreview,PhysRevC.70.031902,DYNDAL2017281,Abbas2013,2017489}. In these collisions, the impact parameter ($b$) is larger than twice the nuclear radium ($R_{A}$).

Recently, significant excesses of J/$\psi$ yield at very low transverse momentum ($p_{T} <$ 0.3 GeV/c) have been observed by the ALICE~\cite{LOW_ALICE} and STAR~\cite{1742-6596-779-1-012039} collaborations in peripheral Hadronic Heavy-Ion Collisions (HHICs). These excesses cannot be explained by the hadronic J/$\psi$ production with currently known cold and hot medium effects taken into account, however, could be qualitatively described by coherent photonuclear production mechanism~\cite{PhysRevC.93.044912,Zha:2017jch,Shi:2017qep}. Assuming that the coherent photonuclear production is response for the observed $J/\psi$ excess, the coherent two-photon process should be there and contribute to the dilepton production in HHICs. Recently, STAR made measurements of $e^{+}e^{-}$ pair production at very low $p_{T}$, and indeed, a significant excess with respect to hadronic cocktails in peripheral HHICs was observed in the preliminary results~\cite{Brandenburg:2017meb}, which points to evidence of coherent photon-photon interactions in HHICs. There are plenty of theoretical calculations on lepton pair production from coherent photon-photon interactions in UPCs~\cite{UPCreview,BAUR20071,BAUR1990786,Klein:2016yzr,PhysRevC.80.044902,Hencken:1995me,0954-3899-15-3-001,PhysRevC.47.2308,Alscher:1996mja}, however, the calculations in HHICs are absent on the market to constrain the origin of the excess at present. In this article, we report calculations of lepton pair ($e^{+}e^{-}$ and $\mu^{+}\mu^{-}$) production from coherent photon-photon interactions in hadronic A+A collisions at RHIC and LHC energies within the STAR and ALICE acceptance. The centrality and pair mass dependence of lepton pair production at $p_{T} < $ 0.15 GeV/c in Au+Au collisions at $\sqrt{s_{NN}} =$ 200 GeV and Pb+Pb collisions at $\sqrt{s_{NN}} =$ 2.76 TeV are presented. We also compare the calculated results with the contributions from hadronic cocktails, QGP thermal radiation, and in-medium $\rho$ broadening in Au+Au collisions at $\sqrt{s_{NN}} =$ 200 GeV within STAR acceptance.

According to the equivalent photon approximation method, a two-photon reaction can be factorized into a semiclassical part and a quantum part. The semiclassical part deals with distribution of massless photons swarming about the colliding ions, while the quantum part usually involves the description of the interaction between the two emitted photons. The cross section to produce a lepton pair with pair mass $W$ can be written as~\cite{Klein:2016yzr}:
    \begin{equation}
    \begin{aligned}
    &\sigma (A + A \rightarrow A + A + l^{+}l^{-})
    \\
    & =\int dk_{1}dk_{2} \frac{n(k_{1})}{k_{1}}\frac{n(k_{2})}{k_{2}}\sigma[\gamma \gamma \rightarrow l^{+}l^{-}(W)],
         \label{equation1}
    \end{aligned}
    \end{equation}
where $k_{1}$ and $k_{2}$ are the two photon energies and $n(k)$ is the photon flux at energy $k$. The two photon energies $k_{1}$ and $k_{2}$ determine the pair mass $W$ and rapidity $y$:
    \begin{equation}
k_{1,2} = \frac{W}{2}e^{\pm y}
         \label{equation1_1}
    \end{equation}
and
   \begin{equation}
    y=\frac{1}{2}\text{ln}\frac{k_{1}}{k_{2}}.
         \label{equation1_2}
    \end{equation}

The photon flux induced by nucleus can be modelled using the Weizs\"acker-Williams method~\cite{KRAUSS1997503}. For the point-like charge distribution, the photon flux is given by the simple formula
    \begin{equation}
    n(k,r) = \frac{d^{3}N}{dkd^{2}r} = \frac{Z^{2}\alpha}{\pi^{2}kr^{2}}x^{2}K_{1}^{2}(x)
         \label{equation2}
    \end{equation}
where $n(k,r)$ is the flux of photons with energy $k$ at distant $r$ from the center of nucleus, $\alpha$ is the electromagnetic coupling constant, $x=kr/\gamma$, and $\gamma$ is lorentz factor. Here, $K_{1}$ is a modified Bessel function. For realistic case, the charge distribution in nucleus should be take into account for the estimation of photon flux. A generic formula for any charge distribution can be written as~\cite{PhysRevC.47.2308}:
   \begin{equation}
  \label{equation3}
  \begin{aligned}
  & n(k,r) = \frac{4Z^{2}\alpha}{k} \bigg | \int \frac{d^{2}q_{\bot}}{(2\pi)^{2}}q_{\bot} \frac{F(q)}{q^{2}} e^{iq_{\bot} \cdot r} \bigg |^{2}
  \\
  & q = (q_{\bot},\frac{k}{\gamma})
  \end{aligned}
  \end{equation}
  where the form factor $F(q)$ is Fourier transform of the charge distribution in nucleus.
 In our calculations, we use two-parameter Fermi distribution (called equivalently Woods-Saxon distribution) for charge distribution in nucleus
 \begin{equation}
  \rho_{A}(r)=\frac{\rho^{0}}{1+\exp[(r-R_{\rm{WS}})/d]}
  \label{equation4}
  \end{equation}
  where the radius $R_{\rm{WS}}$ (Au: 6.38 fm, Pb: 6.62 fm) and skin depth $d$ (Au: 0.535 fm, Pb: 0.546 fm) are based on fits to electron scattering data~\cite{0031-9112-29-7-028,DEJAGER1974479}, and $\rho^{0}$ is the normalization factor.
\renewcommand{\floatpagefraction}{0.75}
\begin{figure*}[htbp]
\includegraphics[keepaspectratio,width=0.3\textwidth]{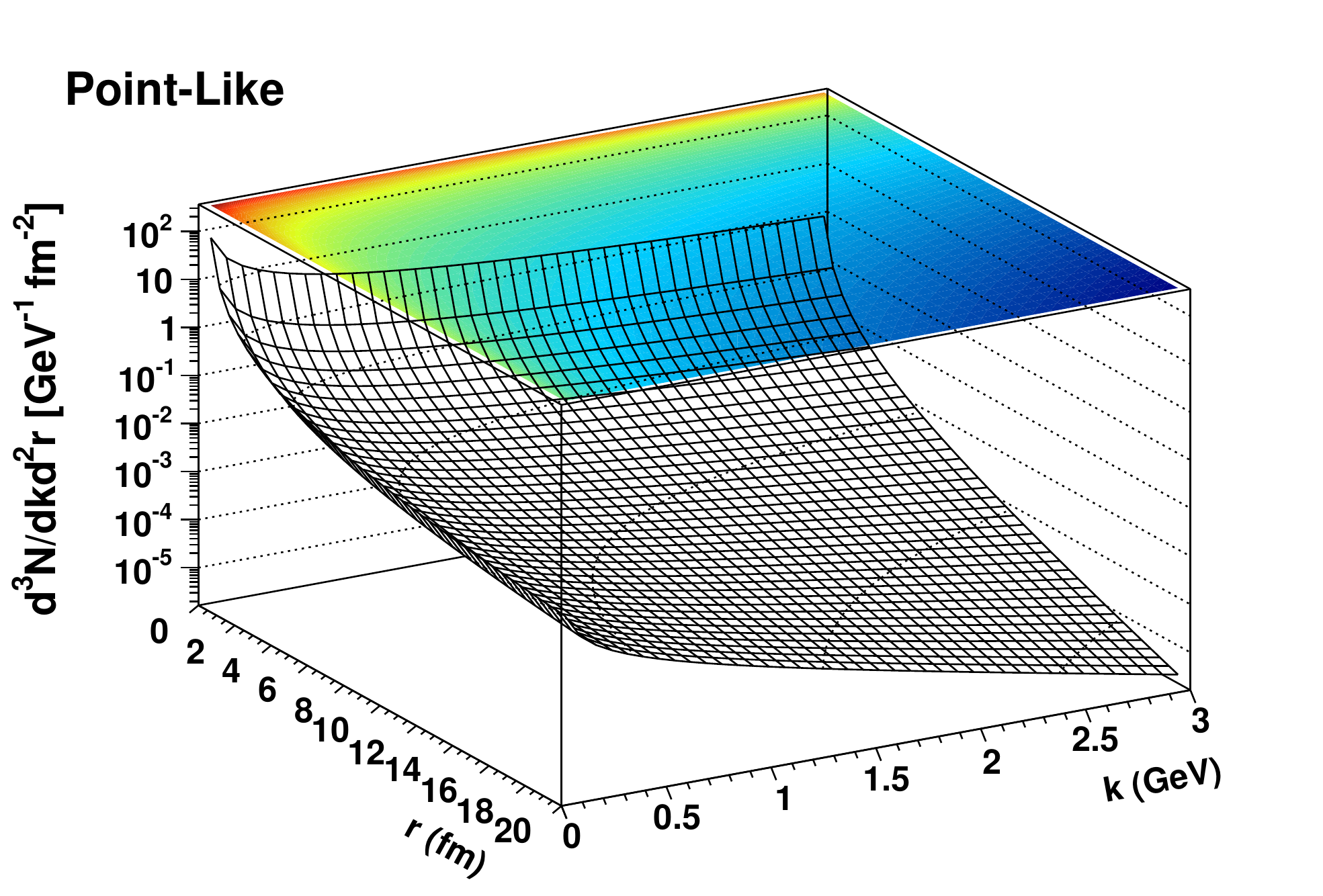}
\includegraphics[keepaspectratio,width=0.3\textwidth]{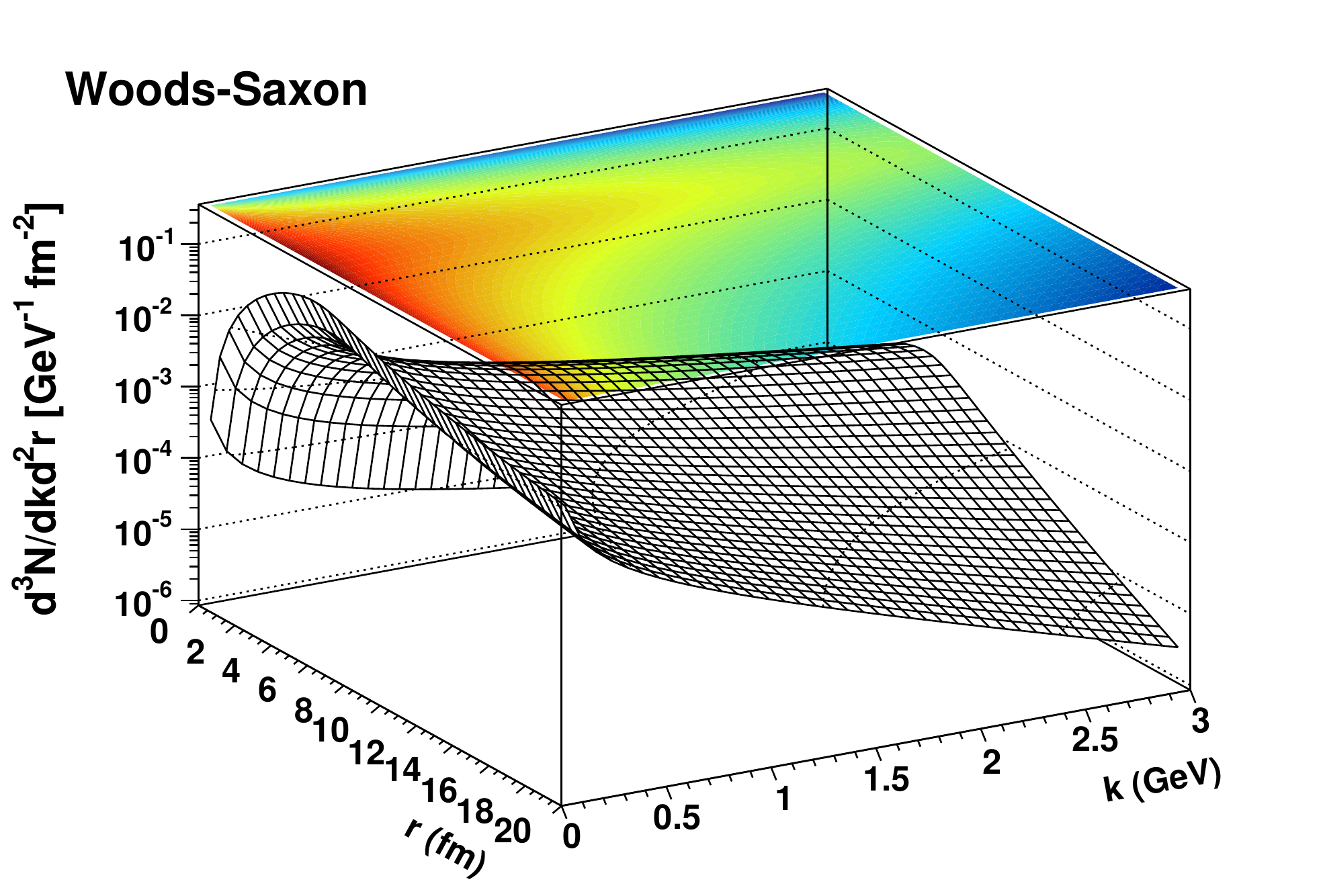}
\includegraphics[keepaspectratio,width=0.3\textwidth]{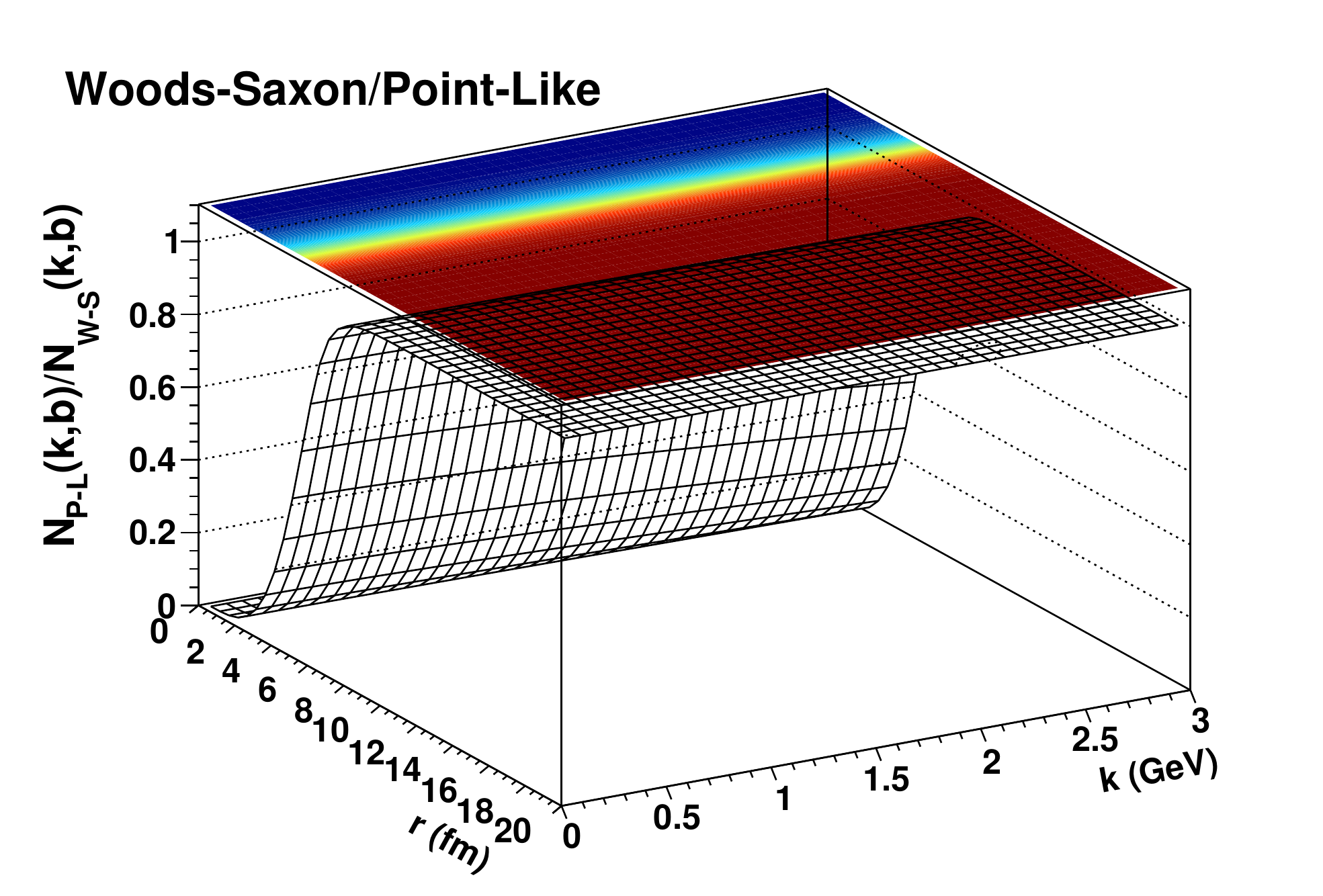}
\caption{Two-dimensional distributions of the photon flux in the distant $r$ and in the energy of photon $k$ for point-like (left panel) and Woods-Saxon (middle panel) form factors in Au+Au collisions at $\sqrt{s_{NN}} =$ 200 GeV. The right panel shows the ratio of the differential photon fluxes with Wood-Saxon form factor to those with point-like form factor.}
\label{figure1}
\end{figure*}

  \renewcommand{\floatpagefraction}{0.75}
\begin{figure*}[htbp]
\includegraphics[keepaspectratio,width=0.7\textwidth]{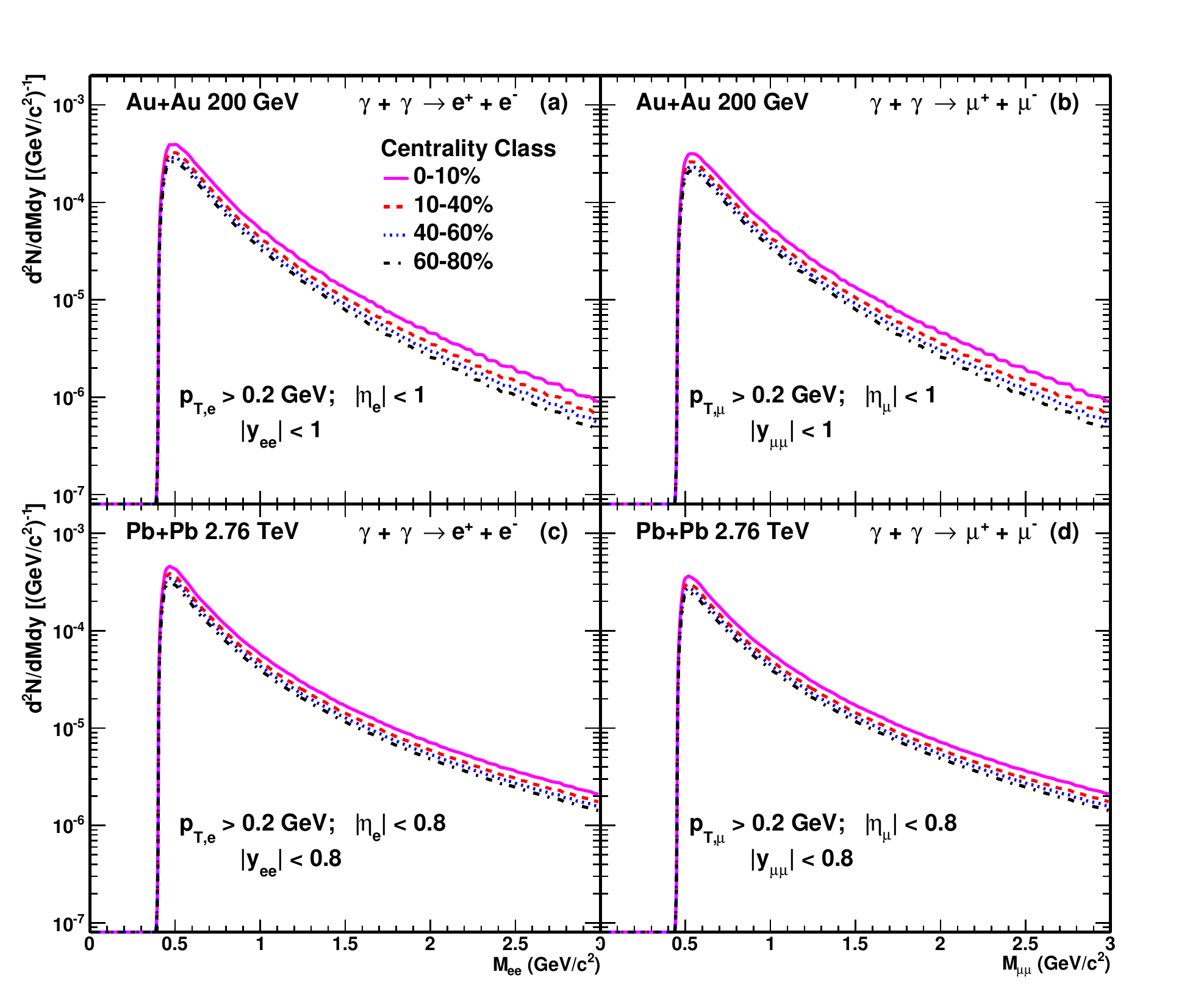}
\caption{The differential pair mass spectrum $d^{2}N/dMdy$ for electron (a) and muon (b) pair with gold beams at RHIC ($\sqrt{s_{NN}} =$ 200 GeV) and for electron (c) and muon (d) pair with lead beams at LHC ($\sqrt{s_{NN}}$ = 2.76 TeV). The different curves in the figure represent the results for different centrality classes. The results are filtered to match the fiducial acceptance described in the text.}
\label{figure4}
\end{figure*}

  \renewcommand{\floatpagefraction}{0.75}
\begin{figure*}[htbp]
\includegraphics[keepaspectratio,width=0.7\textwidth]{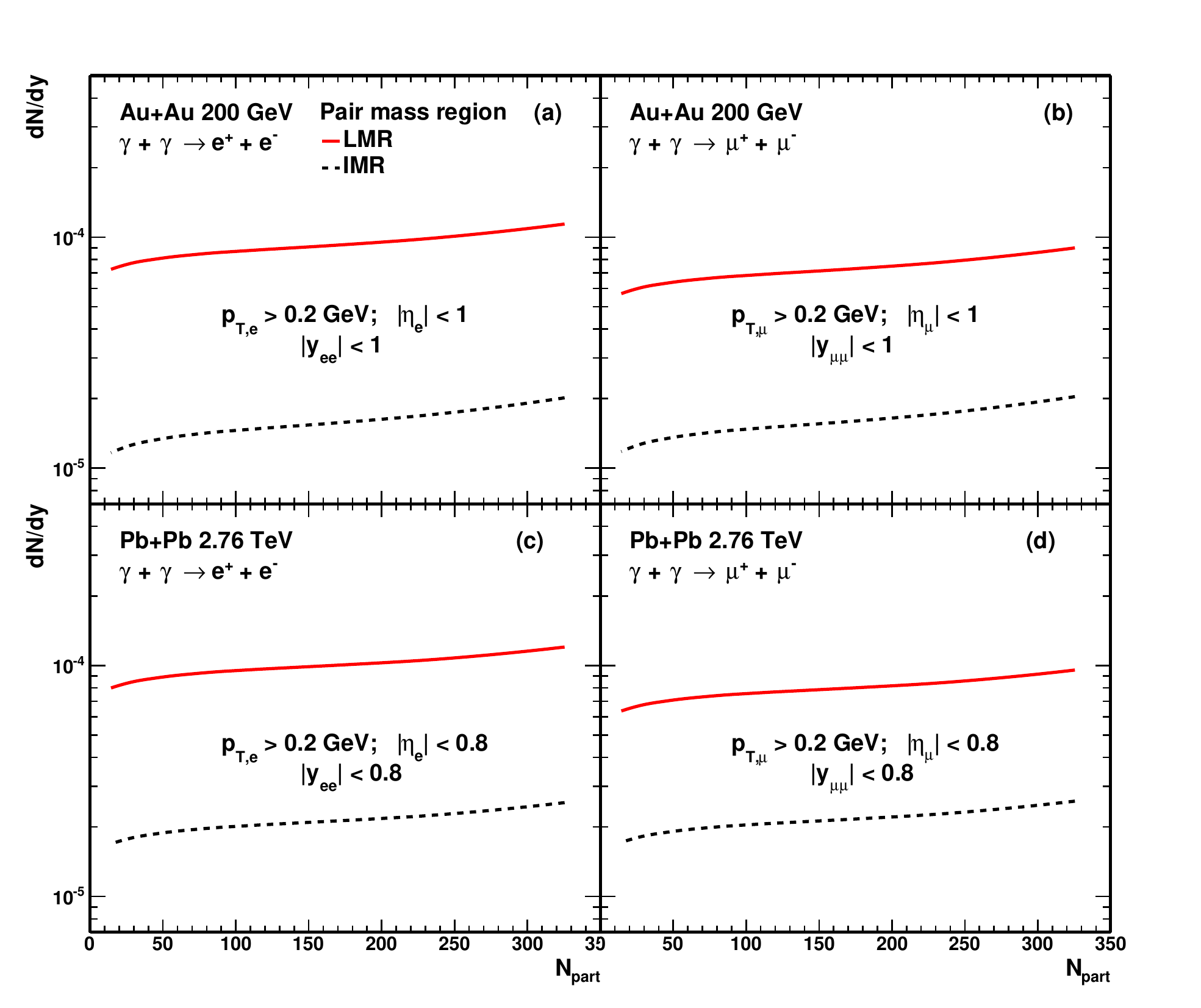}
\caption{The integrated yield $dN/dy$ as a function of $N_{\text{part}}$ for electron (a) and muon (b) pair with gold beams at RHIC and for electron (c) and muon (d) pair with lead beams at LHC. Two curves are shown in each panel: the integrated yield $dN/dy$ for the low mass region (0.4 - 1.0 GeV/c$^{2}$, LMR) and the intermedium mass region (1.0 - 3.0 GeV/c$^{2}$, IMR). The results are filtered to match the fiducial acceptance.}
\label{figure5}
\end{figure*}

  \renewcommand{\floatpagefraction}{0.75}
\begin{figure*}[htbp]
\includegraphics[keepaspectratio,width=0.7\textwidth]{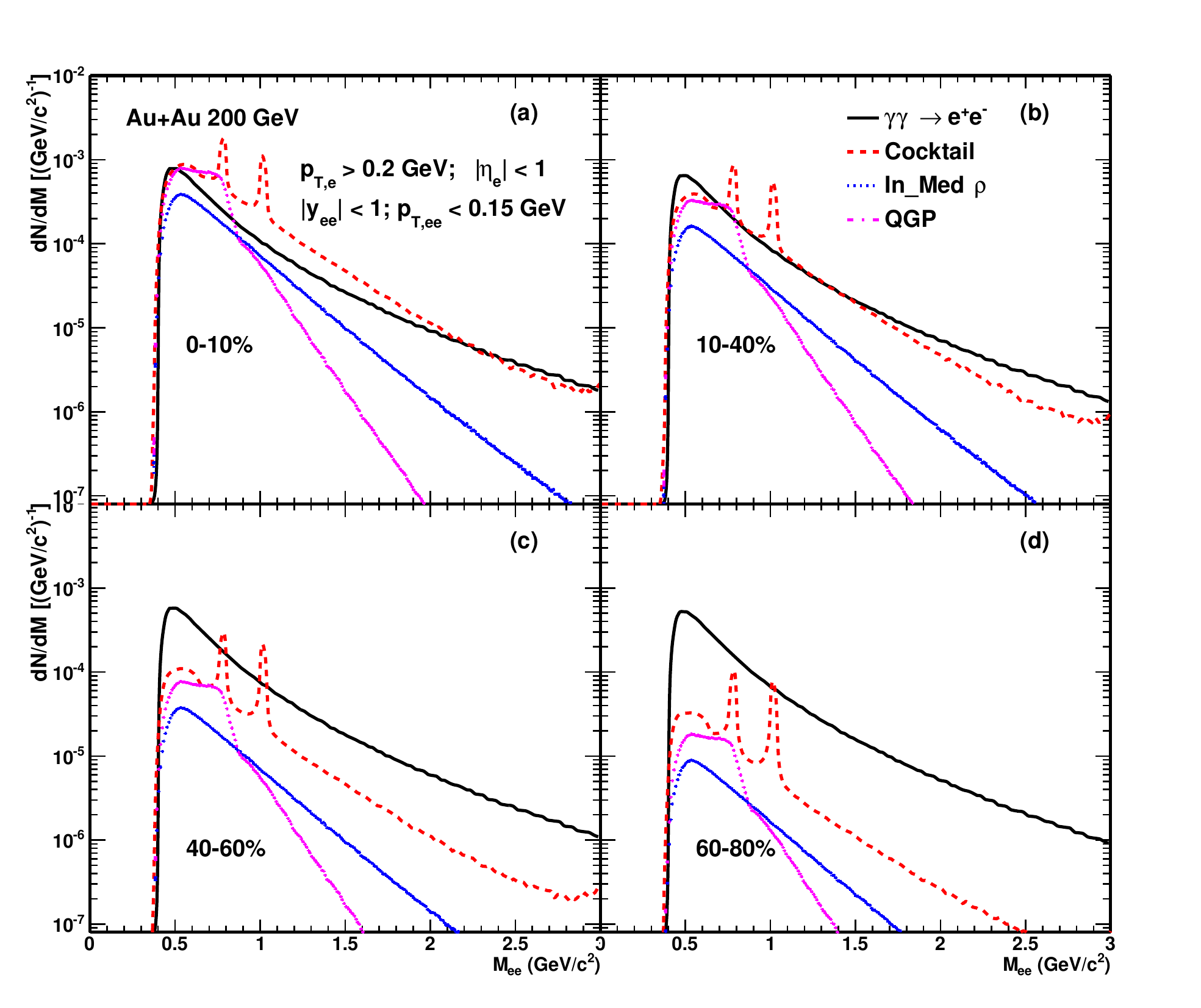}
\caption{The mass spectrum of electron pairs with gold beams at RHIC top energy ($\sqrt{s_{NN}} =$ 200 GeV) for 0-10$\%$ (a), 10-40$\%$(b), 40-60$\%$, and 60-80$\%$ centrality classes. The mass distributions are compared to hadronic cocktail simulations without $\rho$ contribution (dash), in medium $\rho$ mass spectrum (dotted) and QGP thermal radiation (dash-dotted). The in medium $\rho$ mass spectrum includes the contribution from freeze-out $\rho$. The results are filtered to match the fiducial acceptance described in the text.}
\label{figure6}
\end{figure*}
  Figure~\ref{figure1} shows the two-dimensional distributions of the photon flux induced by Au+Au collisions at $\sqrt{s_{NN}} =$ 200 GeV as a function of distant $r$ and energy $k$ for point-like (left panel) and Woods-Saxon (middle panel) form factors. The right panel of Fig.~\ref{figure1} shows the ratio of the differential photon fluxes with Wood-Saxon form factor to those with point-like form factor. One can observe that the difference between photon fluxes obtained with point-like form factor and the result for Woods-Saxon photon source is huge inside the nucleus ($r < R_{WS}$). For $r \gg R_{WS}$, the difference is negligible. In this calculation, for $r < 10$ fm, we use the photon flux calculated with Woods-Saxon form factor; while at  $r > 10$ fm, the result with point-like assumption is employed.

  The elementary cross-section to produce a pair of leptons with lepton mass $m$ and pair invariant mass $W$ can be determined by the Breit-Wheeler formula~\cite{PhysRevD.4.1532}
     \begin{equation}
  \label{equation5}
  \begin{aligned}
  & \sigma (\gamma \gamma \rightarrow l^{+}l^{-}) =
  \\
  &\frac{4\pi \alpha^{2}}{W^{2}} [(2+\frac{8m^{2}}{W^{2}} - \frac{16m^{4}}{W^{4}})\text{ln}(\frac{W+\sqrt{W^{2}-4m^{2}}}{2m})
  \\
  & -\sqrt{1-\frac{4m^{2}}{W^{2}}}(1+\frac{4m^{2}}{W^{2}})].
  \end{aligned}
  \end{equation}
The angular distribution of these lepton pairs is given by
 \begin{equation}
  G(\theta) = 2 + 4(1-\frac{4m^{2}}{W^{2}})\frac{(1-\frac{4m^{2}}{W^{2}})\text{sin}^{2}(\theta)\text{cos}^{2}(\theta)+\frac{4m^{2}}{W^{2}}}{(1-(1-\frac{4m^{2}}{W^{2}})\text{cos}^{2}(\theta))^{2}},
  \label{equation6}
  \end{equation}
 where $\theta$ is the angle between the beam direction and one of the leptons in the lepton-lepton center of mass frame. Here, we neglect the effect of the photon $p_{T}$ on the angular distribution.

 The approach used in this calculation is very similar to that of the STARlight Monte Carlo~\cite{Klein:2016yzr}, which is popular for UPCs. STARlight has been extensively compared with UPC data, with good agreement found for $\gamma + \gamma \rightarrow l^{+} + l^{-}$, with data from STAR~\cite{PhysRevC.70.031902}, ATLAS~\cite{DYNDAL2017281}, ALICE~\cite{Abbas2013} and CMS~\cite{2017489} collaboration. In STARlight, a point-like charge distribution for photon flux is employed and the production for transverse positions within the nucleus is neglected. This approximation is proper for UPCs, because the production inside the nucleus is small. However, comes to the hadronic heavy-ion collision, when the two colliding nuclei are very close to each other, the charge distribution and the production inside the nucleus can not be neglected. In this calculation, we employ the Woods-Saxon charge distribution, and include the production inside the nucleus.

 With the convolution of equivalent photon spectra and elementary cross-section,the probability to produce a lepton pair with pair mass $W$ and rapidity $y$ for a collision at impact parameter $b$ can be given by:
  \begin{equation}
  P(W,y,b)=\frac{W}{2}\int d^{2}r_{1}n(k_{1},r_{1})n(k_{2},|\vec{b} - \vec{r}_{1}|)\sigma_{\gamma \gamma}(W).
  \label{equation7}
  \end{equation}
  The invariant yield for coherent lepton pair production in a selected centrality class can be written as:

\begin{equation}
  \frac{d^{2}N}{dWdy}=\frac{\int^{b_{max}}_{b_{min}} d^{2}b P(W,y,b) \times P_{H}(\vec{b})}{\int^{b_{max}}_{b_{min}} d^{2}b P_{H}(\vec{b})},
  \label{equation16}
\end{equation}
  with the hadronic interaction probability
      \begin{equation}
  P_{H}(\vec{b}) = 1 - \text{exp}[-\sigma_{\text{NN}} \int d^{2}r T_{A}(\vec{r})T_{A}(\vec{r}-\vec{b})]
  \label{equation9}
  \end{equation}
  where $\sigma_{\text{NN}}$ is the inelastic hadronic interaction cross section, $T_{A}(\vec{r})$ is the nuclear thickness function defined as
        \begin{equation}
T_{A}(\vec{r}) = \int dz \rho(\vec{r},z)dz.
  \label{equation10}
  \end{equation}

 The $b_{min}$ and $b_{max}$ in Eq.~\ref{equation16} are the minimum and maximum impact parameters for a given centrality bin, and can be determined from the Glauber Monte Carlo simulations~\cite{Loizides:2014vua}. To estimate the uncertainties of the calculation, the Radius $R_{\text{WS}}$ and skin depth $d$ in Eq.~\ref{equation4} have been varied by a factor of 2$\%$ and 10$\%$, respectively. The extracted relative uncertainty is about 8$\%$.

 In hadronic heavy-ion collisions, the coherent lepton pair production is accompanied by violent hadronic interactions occurred in the overlap region, which may impose impact on the coherent production. The possible disruptive effects can be considered into two part: photon emission and production in the overlap region. For photon emission, the photon field travels along with the colliding nucleus and they are likely to be emitted before the hadronic interactions occur by about $\Delta t \sim r/c$, where r is the distant from the nucleus. The energetic hadronic collisions in the overlap region happen at a small time scale $t_{\text{coll}} \sim R_{WS}/\gamma c$. Therefore, due to the retarded time effect, the photon emission should be unaffected by hadronic interactions. Furthermore, the photon bremsstrahlung from the nucleons in the overlap region~\cite{PhysRevD.31.63,PhysRevC.33.153} can also contribute to the lepton pair production. The production rate is proportional to $Z^{4}_{\text{eff}}$, where $Z_{\text{eff}}$ is the charge in overlap region for one nucleus. In contrast to the production from virtual photons ,which is proportional to $Z^{4}$, the contribution from bremsstrahlung should be small, at least for peripheral collisions ($Z_{\text{eff}} << Z$). For the coherent production in the overlap region, the product may be affected by the violent interactions, leading to the destruction of coherent action. To elaborate this effect, we make calculations without production in the overlap region, it brings down the cross section by about 20$\%$. Because the final product is lepton pair, which is not subject to the strong interactions in the overlap region, the destruction effect should be small. In this calculations, we neglect the possible disruption from hadronic interactions. This is appropriate for peripheral collisions, however, it may be questionable in central collisions, which needs more robust consideration in future work.

The cross-section is heavily concentrated among near-threshold pairs, which are not visible to existing detectors. So, calculations of the total cross-section are not particularly useful. Instead, we calculate the cross-section and kinematic distributions for two samples, with acceptances that match those used by STAR and ALICE.

Figure~\ref{figure4} shows the differential pair mass spectrum $d^{2}N/dMdy$ for electron (a) and muon (b) pair with gold beams at RHIC ($\sqrt{s_{NN}} =$ 200 GeV) and for electron (c) and muon (d) pair with lead beams at LHC ($\sqrt{s_{NN}} =$ 2.76 TeV). The different curves in the figure represent calculations for different centrality classes. The results are filtered to match the fiducial acceptance: daughter track transverse momentum $p_{T} >$ 0.2 GeV/c, track pseudo-rapidity $|\eta| <$ 1, and pair rapidity $|y| <$ 1 at RHIC; daughter track $p_{T} >$ 0.2 GeV/c, track $|\eta| <$ 0.8, and pair rapidity $|y| < $ 0.8 at LHC. The mass spectrums for both electron and muon pair possess little centrality dependence. The production of electron pair in the fiducial acceptance is slightly higher than that of muon pair at low mass region, however, the difference becomes negligible when comes to higher mass region. By comparison with the calculations at RHIC and LHC, one can observe that the mass spectrum at LHC is harder than that at RHIC.

Figure~\ref{figure5} shows the integrated yield $dN/dy$ as a function of number of participate nucleons ($N_{\text{part}}$) for electron (a) and muon (b) pair with gold beams at RHIC and for electron (c) and muon (d) pair with lead beams at LHC. Two curves are shown in each panel: the integrated yield $dN/dy$ for the low mass region (0.4 - 1.0 GeV/c$^{2}$, LMR) and the intermedium mass region (1.0 - 3.0 GeV/c$^{2}$, IMR). The results are filtered to match the fiducial acceptance. The integrated yield within the fiducial acceptance shows weak centrality dependence. The production rate at LHC is slightly higher than that at RHIC for low mass region (LMR), and the difference becomes significant for intermedium mass region (IMR).

In hadronic heavy-ion collisions, the violent strong interactions can also contribute to the lepton pair mass spectrum, which may overwhelm the signals from coherent production. Figure~\ref{figure6} shows the mass spectrum of electron pairs with gold beams at RHIC top energy ($\sqrt{s_{NN}} =$ 200 GeV) for 0-10$\%$ (a), 10-40$\%$(b), 40-60$\%$, and 60-80$\%$ centrality classes, compared to hadronic cocktail simulations without $\rho$ contribution (dash line), in medium $\rho$ mass spectrum (dotted line) and QGP thermal radiation (dash-dotted line). The results are filtered to match the fiducial acceptance: daughter track $p_{T} >$ 0.2 GeV/c, track $|\eta| <$ 1, pair rapidity $|y| < $1, and pair $p_{T} < $ 0.15 GeV/c. The hadronic cocktail is a Monte Carlo (MC) simulation performed to account for the contributions from known hadronic sources, which include the decays or Dalitz decays of $\pi^{0}$, $\eta$, $\eta^{\prime}$, $\omega$, $\phi$, $J/\psi$ and $\psi^{\prime}$. The dielectron yields from correlated charm or bottom decays and Drell-Yan production are based on PYTHIA model~\cite{SJOSTRAND2001238} calculations scaled by number of binary collisions in which parameters have been tuned to published STAR measurements~\cite{PhysRevLett.113.142301}. The procedure is the same as Ref.~\cite{PhysRevLett.113.022301,Adamczyk201564,2017451}. The QGP thermal radiation and the broadened $\rho$ mass spectrum are from an effective many-body model~\cite{PhysRevLett.97.102301,VANHEES2008339,PhysRevC.63.054907,Rapp:2013nxa}, which has successfully explained the SPS and RHIC data. The in medium $\rho$ mass spectrum includes the contribution from freeze-out $\rho$. For central collisions (0-10$\%$), the hadronic contribution is dominant over the coherent production at relative low mass region(0.4-2 GeV/c$^{2})$, but they come close to each other at higher mass region. When comes to semi-central (10-40$\%$) collisions, the hadronic background for continuum region is comparable to that of coherent electron pair production. Strikingly, for peripheral collisions (40-80$\%$), one can find that the coherent electron pair production is dominant over the contributions from cocktail, QGP thermal radiation and the in-medium $\rho$. In comparison with STAR measurements~\cite{Brandenburg:2017meb}, the calculations could reasonably describe the excess observed at STAR, which points to evidence of coherent photon-photon interaction in HHICs.

In summary, we have preformed calculations of lepton pair ($e^{+}e^{-}$ and $\mu^{+}\mu^{-}$) production from coherent photon-photon interactions in hadronic A+A collisions at RHIC and LHC energies within the STAR and ALICE acceptance. The centrality and pair mass dependence of lepton pair production at $p_{T} < $ 0.15 GeV/c in Au+Au collisions at $\sqrt{s_{NN}} =$ 200 GeV and Pb+Pb collisions at $\sqrt{s_{NN}} =$ 2.76 TeV are studied. We also compare the calculated results with the contributions from hadronic cocktails, QGP thermal radiation, and in-medium $\rho$ broadening in Au+Au collisions at $\sqrt{s_{NN}} =$ 200 GeV within STAR acceptance and find that the coherent electron pair is dominant over the hadronic contributions for peripheral collisions, which may account for the significant excess observed at STAR. In present calculations, the possible disruption from violent hadronic interactions is neglected, which needs more robust consideration in future work.  This work calls for further experimental and theoretical efforts to confirm the coherent lepton pair production in violent hadronic A+A collisions.

We thank Prof. Spencer Klein and Pengfei Zhuang for useful discussions. This work was funded by the National Natural Science Foundation of China under Grant Nos. 11505180, 11775213 and 11375172, the U.S. DOE Office of Science under contract No. DE-SC0012704, and MOST under Grant No. 2014CB845400.

\nocite{*}
\bibliographystyle{aipnum4-1}
\bibliography{aps}
\end{document}